\documentclass[fleqn,usenatbib]{mnras}

\usepackage{newtxtext,newtxmath}

\usepackage[T1]{fontenc}

\DeclareRobustCommand{\VAN}[3]{#2}
\let\VANthebibliography\thebibliography
\def\thebibliography{\DeclareRobustCommand{\VAN}[3]{##3}\VANthebibliography}

\usepackage{graphicx}	
\usepackage{amsmath}	
\usepackage{multicol,multirow}
\usepackage{caption}
\usepackage{tabularx}
\usepackage{color,soul}
\usepackage{cleveref}

\newcommand{\nustar}{{\it NuSTAR}}

\newcommand{\ms}{$M_{\odot}$ }

\newcommand{\ixpe}{\textit{IXPE}}
\newcommand{\monk}{\textsc{monk }}

\newcommand{\ergs}{\,erg/s}
\newcommand{\degree}{^\circ}

\title[Polarization properties of WM-NSs in LMXBs]{Polarization properties of weakly magnetized Neutron Stars in Low Mass X-ray Binaries}

\author[Gnarini et al.]{
A. Gnarini,$^{1}$\thanks{E-mail: andrea.gnarini@uniroma3.it}
F. Ursini,$^{1}$
G. Matt,$^{1}$
S. Bianchi,$^{1}$
F. Capitanio,$^{2}$
M. Cocchi,$^{2}$
R. Farinelli,$^{3}$
and W. Zhang$^{4}$
\\
$^{1}$Dipartimento di Matematica e Fisica, Università degli Studi Roma Tre, via della Vasca Navale 84, 00146 Roma, Italy\\
$^{2}$INAF, IAPS, via del Fosso del Cavaliere 100, I-00113 Roma, Italy \\
$^{3}$INAF - Osservatorio di Astrofisica e Scienza dello Spazio di Bologna, Via P. Gobetti 101, I-40129 Bologna, Italy \\
$^{4}$National Astronomical Observatories, Chinese Academy of Sciences, 20A Datun Road, Beĳing 100101, China \\
}

\date{Accepted XXX. Received YYY; in original form ZZZ}

\pubyear{2022}

\begin{document}
\label{firstpage}
\pagerange{\pageref{firstpage}--\pageref{lastpage}}
\maketitle

\begin{abstract}
X-ray polarimetry missions like \ixpe \ will be able to measure for the first time the polarization properties of accreting, weakly magnetized neutron stars in Low Mass X-ray Binaries. In this work we present simulations of the expected X-ray polarized signal including the coronal emission for different geometries of the corona itself, i.e. a slab above the accretion disc and a spherical shell around the neutron star. The simulations are performed with the fully relativistic Monte Carlo code \textsc{monk} capable of computing the X-ray polarization degree and angle for various physical input parameters of the neutron star, disc and corona. Different coronal geometries result in significantly different X-ray polarization properties, which can therefore be used to constrain the geometry of the systems.
\end{abstract}




\section{Introduction}


Accreting, weakly magnetized Neutron Stars in Low-mass X-ray binaries (NS-LMXBs) are amongst the brightest X-ray sources in the sky, with luminosities ranging in the $10^{36}-10^{38}$ \ergs~interval. They accrete matter via Roche-lobe overflow from a stellar companion, typically a late Main Sequence star with a mass generally lower than the solar mass or, not uncommonly, from a dwarf evolved object. 
They are very variable emitters, even during a single observation.

NS-LMXBs are traditionally classified in a few broad families by their joint timing and spectral properties as observed in the ``classic" X-ray band, 1–10 keV.  
Following their tracks in the X-ray Hard–colour/Soft–colour diagram (CCD) and their CCD position correlated timing behavior \cite[quasi-periodic oscillations, low-frequency noise,][]{VanDerKlis.1989}, their standard classification \citep{Hasinger.VanDerKlis.1989} includes:
\begin{itemize}
    \item High Soft State (HSS) Z sources ($> 10^{38}$ \ergs);
    \item HSS bright Atoll sources ($10^{37}-10^{38}$ \ergs);
    \item Low Hard State (LHS) Atoll sources ($\sim 10^{36}$ \ergs).
\end{itemize}

The Z sources exhibit a wide Z-like three-branch pattern in the CCD, while Atolls show less extended, more compact tracks in a rounded single spot in the hard region of the CCD (LHS atolls), termed ``island" state or, for bright atolls, in a ``banana" shape \citep{VanDerKlis.1995}.  
The majority of persistent NS-LMXBs are generally observed either in HSS or (less frequently) in LHS, but most of the transients and several persistent sources can easily perform state transitions from LH to HS and vice-versa in a relatively short timescale. 
The spectral and timing properties of weakly magnetized NS-LMXBs are very similar to those  observed in black hole LMXBs \citep[see][for more details]{Munoz-Darias.2014,Motta.etAl.2017}.
The evolution of the involved physical parameters (plasma temperature, accretion rate, inner disc radius, etc) define the characteristics of the spectral states of NS-LMXBs. Corona temperature and, consequently, its opacity clearly distinguish LHS from HSS sources. LHS coronae are much hotter and transparent ($kT_e \gtrsim 20$ keV, $\tau \sim 1-2$) with respect to the HSS ones ($kT_e \sim 3$ keV, $\tau \gtrsim 4-5$ depending on the geometry of the corona itself). Intermediate states with plasma temperature $\sim 10$ keV are rarely observed as the LH-HS state transition timescale is generally fast and not predictable \citep[see e.g.][for a lucky case]{Marino.etAl.2019}.

The X-ray emission of NS-LMXBs is generally modelled with two main spectral components, each of them connected to a distinct soft photon population: a thermal emission, i.e. a black-body (either from the NS surface or the boundary layer) or a multicolour disc emission, characterising the spectral shape below $\sim 10$ keV, and a harder component related to inverse Compton scattering of these soft photons by the hot electron plasma in the corona.
In the traditional \textit{eastern-like} scenario \citep{Mitsuda.etAl.1984,Mitsuda.etAl.1989} the cold photon population is described by a multicolour disc black-body and is directly observed, while the hotter NS population is (almost) completely Comptonised by the hot electron plasma. The opposite occurs in the \textit{western model} \citep{White.etAl.1988}: the hotter, NS population is directly detected and modelled by pure black-body while the cold, disc photons are Compton-upscattered by the electron corona. Both the models deal with two separate black-body (or multicolour) distributed photon populations and a single Comptonising region.

Despite the two models being proposed more than 30 years ago, it is still a matter of debate what is the right one, because they are largely degenerate spectroscopically. It might even be possible that both seed photon populations get scattered in the corona \citep{Cocchi.etAl.2011}.
Moreover, the geometry of the corona, and therefore its origin, is still uncertain. 
However, frequency-resolved spectra of the atoll source 4U 1608 at high luminosity in banana state by \cite{Gilfanov.etAl.2003} have shown that the lower kHz QPO has the same spectral shape as the boundary layer (i.e. Comptonized) component. 
Since the most rapid variability is to be associated to the boundary layer rather than to the disc, this strongly points in favour of a Eastern-like scenario for the accreting geometry of the sources \citep[see also review by][]{Done.etAl.2007}.
The frequent observation of a Fe emission line at $\sim 6$ keV, especially in HSS sources, strongly suggests that Compton reflection by a colder medium (as the outer accretion disc itself) is a further spectral component to be taken into account.

For the first time thanks to the NASA/ASI Imaging X-ray Polarimetry Explorer \citep[\ixpe,][]{Weisskopf.etAl.2016,Weisskopf.2022}, it might be possible to measure the polarimetric properties of the X-ray radiation coming from NS-LMXBs.
\ixpe \ operates in the 2-8 keV band \citep[][]{Weisskopf.2022} and is equipped with three X-ray telescopes with polarization-sensitive imaging detectors \citep[the gas-pixel detectors,][]{Costa.2001}.
\ixpe \ was successfully launched on December 9, 2021 and in the first year observing plan will include observations of bright atoll/Z sources, namely GS 1826-238, Cygnus X-2 and GX 9+9.
Combining spectral and polarimetric observations, it might be possible to constrain several different physical parameters of the NS, the disc and the corona, especially trying to understand the shape and the dimensions of the corona itself.

In this work, we present numerical simulations of the X-ray polarized radiation from NS-LMXBs, including coronal emission with the use of the general relativistic Monte Carlo code \textsc{monk} \citep[][submitted]{Zhang.etAl.2019,Zhang.etAl.2022} suitably adapted for NS-LMXBs, including also photons emitted from the NS surface. 

The paper is structured as follows. We discuss the numerical setup, reviewing the basics and the capabilities of \textsc{monk} and exploring different configurations of the corona in Sect. \ref{sec:Simulations}.
We present then the results of the simulations assuming or not intrinsic polarization and for different observer's viewing inclinations in Sect. \ref{sec:Results}. We compare then the polarization degree and angle behaviour for the two chosen geometries and for NS-LMXBs in HSS or LHS. Finally, we summarize our conclusions in Sect. \ref{sec:Conclusions}.  

\section{Numerical Setup}\label{sec:Simulations}


\monk is a general relativistic Monte Carlo radiative transfer code capable of calculating the spectral and polarization properties of Comptonized radiation coming from a corona illuminated by both the NS and a standard accretion disc \citep{Novikov.Thorne.1973}. 

For a NS-LMXB, seed photons emitted by the neutron star surface have to be considered in addition to those of the disc. A black-body spectrum with $kT_\text{ns}$ is assumed to model the unpolarized neutron star surface emission.
The seed photons from the disc are generated instead according to the disc emissivity, with the option to assume either unpolarized radiation, or an initial polarization given by \cite{Chandrasekhar.1960} for a semi-infinite plane-parallel atmosphere. 
Once emitted, the seed photons are ray-traced along null geodesics in Kerr spacetime, transporting the polarization vector. When a photon reaches the corona is Compton scattered, assuming the Klein-Nishina cross section. 
The Stokes parameters of the scattered photons are computed in the electron rest frame \citep{Connors.etAl.1980} and then transformed in the observer (Boyer-Lindquist) frame. The propagation terminates when the photons either hit the star surface, the disc or arrive at infinity. 
The energy and polarization spectrum is produced by counting the photons arrived to the observer at infinity.  Since the radiation produced by Compton scattering is linearly polarized, the Stokes parameter $V$ is set at zero, with the code computing only $I$, $Q$ and $U$. The polarization degree and angle are then, as customary, defined as
\begin{equation}
    \label{eq:Pol.Degree}
    \delta = \frac{\sqrt{Q^2 + U^2}}{I} \, ,
\end{equation}
\begin{equation}
    \label{eq:Pol.Angle}
    \psi = \frac{1}{2} \arctan \bigg( \frac{U}{Q} \bigg) \, .
\end{equation}

\monk requires several different input parameters: the physical parameters of the neutron star (mass, radius and period); the accretion rate and the disc parameters; the optical depth $\tau$, the temperature $kT_e$ and the geometrical parameters of the Comptonizing region. 

For all simulations, we consider a standard neutron star with $M = 1.4$ $M_\odot$, $R = 12$ km. Regarding the NS period, we assume $P = 3$ ms in analogy to the one derived from QPOs by \cite{Wijnands.etAl.1998} for Cygnus X-2 \citep[see also][for a statistical analysis of the spin distributions of NS-LMXBs]{Patruno.2017}.
Since photons propagate along null geodesics in Kerr spacetime, the spin parameter $a$ is required. For a standard neutron star, we can use the period in order to derive the spin, adopting $a=0.47/P(\mathrm{ms}) = 0.1567$ \citep[see][]{Braje.etAl.2000}. 
The temperature of the neutron star is $kT_\text{ns} = 1.5$ keV, both for Atolls and Z-sources \citep[see \citealt{Rev.Gilfanov.2006}, for more details; ][]{Farinelli.etAl.2008,GX9+9.2020}.

Some of the disc input parameters, such as the accretion rate $\dot{M}$ and the inner disc radius $R_\text{in}$, depend on the spectral state of the NS-LMXB.
The disc can either extend down to the boundary layer or truncated at an arbitrary radius. The disc cannot extend indeed all the way to the neutron star surface, because of the magnetic field which stops the accreting plasma at a position where the magnetic pressure and the plasma one become of the same order. Then the accreted matter starts flowing toward the neutron star along the magnetic field lines. 
Depending on the NS-LMXBs spectral state, we have considered typical values for the accretion rate $\dot{M}$ \citep{Paizis.etAl.2016} in terms of the Eddington accretion rate $\dot{M}_\text{Edd}$ for a 1.4 \ms star as
\begin{equation}
\label{eq:Acc.Rate}
    \dot{M} = \frac{4 \pi G m_p M}{c \, \sigma_T} \, \dot{m} = \dot{M}_\text{Edd} \, \dot{m} \approx 1.93 \times 10^{17} \, \dot{m} \, \mathrm{g \, s^{-1}}
\end{equation}
where $\dot{m}$ is the adimensional accretion rate.

\begin{figure}
    \centering
    \includegraphics[width=0.45\textwidth]{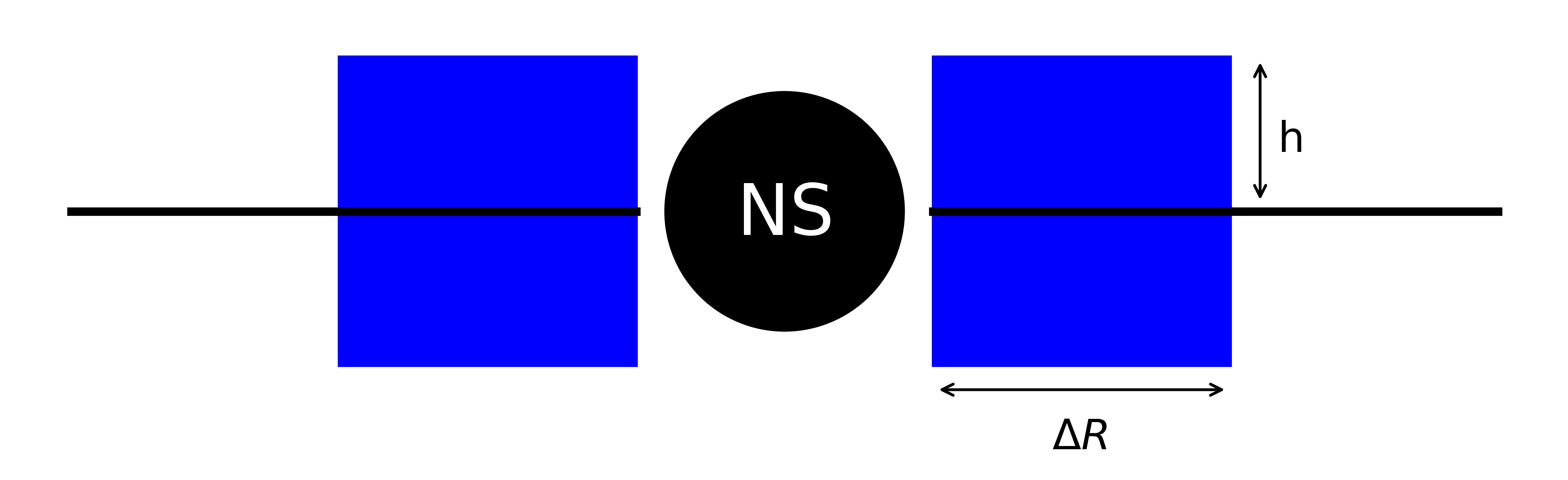}
    \includegraphics[width=0.45\textwidth]{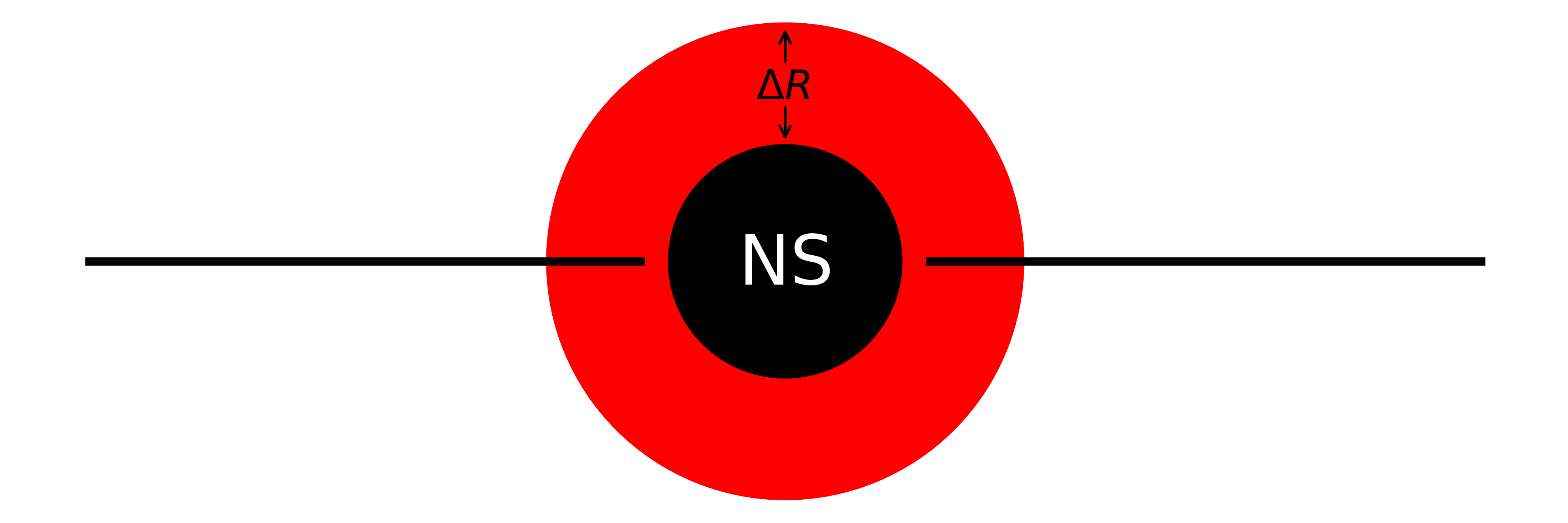}
    \caption{Schematic representation of the different coronal geometries: the slab (upper panel) and the spherical shell (lower panel).}
    \label{fig:Corona.Geometries}
\end{figure}

\captionsetup[table]{belowskip=0pt} 

\begin{table*}
\centering
\caption{Physical and geometrical parameters respectively of the disc and corona used as input for \textsc{monk} simulations. We consider a standard NS with $M = 1.4$ $M_\odot$, $R=12$ km and $P=3$ ms for all the simulations. The accretion rate is shown in terms of the adimensional accretion rate $\dot{m}$ (see Eq. \ref{eq:Acc.Rate}). The optical depth for the spherical shell is radial, while for the slab is vertical and measured from the half-plane of the disc.}
\begin{tabular}{cccccccc}
\multicolumn{1}{c}{} & \multicolumn{3}{c}{Disc Parameters} & \multicolumn{4}{c}{Corona Parameters} \\ 
\hline
\hline
Shape & $\dot{m}$ & $R_\text{in}$ & $f_\text{col}$ & \multicolumn{2}{c}{Geometrical Parameters} & $kT_e$ & $\tau$ \\
\hline
Slab (HSS) & 1 & 7 $R_\text{G}$ & 1.6 & $\Delta R = 15 \, R_\text{G}$ & $H = 7.5 \, R_\text{G}$ & 3 keV & 4 \\
Shell (HSS) & 1 & 7 $R_\text{G}$ & 1.6 & \multicolumn{2}{c}{$\Delta R = 6 \, R_\text{G}$} & 3 keV & 8 \\
Slab (LHS) & 0.05 & 6 $R_\text{G}$ & 1.6 & $\Delta R = 15 \, R_\text{G}$ & $H = 7.5 \, R_\text{G}$ & 25 keV & 1 \\
Shell (LHS) & 0.05 & 6 $R_\text{G}$ & 1.6 & \multicolumn{2}{c}{$\Delta R = 6 \, R_\text{G}$} & 25 keV & 2 \\
\hline
\end{tabular}
\label{tab:Sim}
\end{table*}

Unless explicitely stated otherwise, the disc emission is assumed to be polarized. As customary, we assume the Chandrasekhar (1960) prescription, valid for a plane-parallel, semi-infinite, pure scattering atmosphere. The polarization degree ranges from zero, when the disc is seen face-on, to almost 12\%, when the disc is seen edge-on. The effects of assuming no intrinsic poalrization will be briefly discussed at the end of section 3. 

The HSS is characterized by a high mass accretion rate very close to the critical (Eddington) values \citep{Farinelli.etAl.2008,CygX2.2009}. For LHS sources, instead, the mass accretion rate is relatively low \citep{Falanga.etAl.2006}. 
The color correction factor varies with the spectral state \citep[e.g.][]{Merloni.etAl.2000}, however slight differences will not qualitatively affect the results.
The color correction factor is set to $f_\text{col} = 1.6$ for both spectral states and geometries \citep{Shimura.Takahara.1995}. 
See Table \ref{tab:Sim} for all the disc and corona parameters.

The physical parameters of the corona, $kT_e$ and $\tau$, are fixed according to the NS-LMXBs spectral state. HSS-LMXBs are characterized by lower temperatures of the Comptonizing region, $kT_{e,\text{HSS}} = 3$ keV, while for LHS sources the corona is much hotter, $kT_{e,\text{LHS}} = 25$ keV; moreover, HSS sources present a significantly more opaque corona, i.e. $\tau_\text{HSS} = 8$, with respect to LHS corona which is more transparent, $\tau_\text{LHS} = 2$ \citep{CygX2.2009,GS1826.2010,Cocchi.etAl.2011,GS1826.2020}. 
These values refer to a spherical corona for which $\tau$ is the radial one. However, in \textsc{monk}, the optical depth for a slab geometry is defined as $\tau = n_e \sigma_T h$ where $h$ is the half-thickness of the slab \citep[\citealt{Zhang.etAl.2019}, see also][for the same definition]{Titarchuk.1994,Poutanen.Svensson.1996}. 
Therefore, for the slab we assumed a value of $\tau$ a factor of 2 smaller than the one of the spherical shell, to obtain a similar spectral shape \cite[see][for more details]{Ursini.etAl.2022}. 

Different geometries of the corona can be explored with \monk \citep{Zhang.etAl.2019}. We considered two different configurations: the slab and the spherical shell (Figure \ref{fig:Corona.Geometries}).
The spherical shell configuration can be considered as representative of the Eastern model, since most of the NS seed photons get scattered and Comptonised by the hot corona surrounding the NS while almost the entire disc can be directly observed (the shell covers only a small part of the disc and only a few photons emitted by the disc are scattered). 
On the other hand, the slab geometry is more representative of the Western model, because most of the disc photons are scattered by the corona, while the NS radiation can be directly detected, depending on the inclination (since the thickness of the slab is comparable to the NS radius).

Therefore, for both geometries, photons coming from both the NS and the disc are at least partly scattered by the electron plasma in the corona.  
With \textsc{monk} both scattered contributions are considered together, along with the direct radiation coming from the NS surface and the disc. Depending on the geometry of the corona and the viewing inclination, the relative fluxes of the different components are then fixed (see Figures \ref{fig:Slab.Data} and \ref{fig:Shell.Data}, lower panels).

The physical parameter of the NS, disc and corona are general reference values for HSS and LHS NS-LMXBs chosen in order to obtain spectra similar to those observed for different LMXBs sources. 
The spectra and the results for the polarization obtained with this set of parameters, therefore, do not refer to any particular source. 
A complete analysis of observed NS-LMXBs will have to be done individually for each source combining spectral and polarimetric observation and using the best-fit spectral parameters as input values for \textsc{monk}.
However, we have verified that the spectra reproduced by \textsc{monk} are in good agreement with the \nustar \ spectrum of GS 1826-238 \citep[see][]{Chenevez.etAl.2016}.

\subsection{Slab}\label{sec:Slab}

The slab corona is assumed to cover part of the disc, starting from the inner disc radius $R_\text{in}$ and extending for $\Delta R = 15 \, R_\text{G}$, in units of gravitational radii ($R_\text{G} = GM_\text{NS}/c^2$), and to co-rotate with the Keplerian disc \citep{Zhang.etAl.2019}.
The vertical thickness $H$ of the slab corona is set to $7.5 \, R_\text{G}$, in order to cover most of the neutron star surface. 
Both the geometrical and physical parameters of the disc and the corona are summarized in Table \ref{tab:Sim}. 

\begin{figure}
    \centering
    \includegraphics[width=0.45\textwidth]{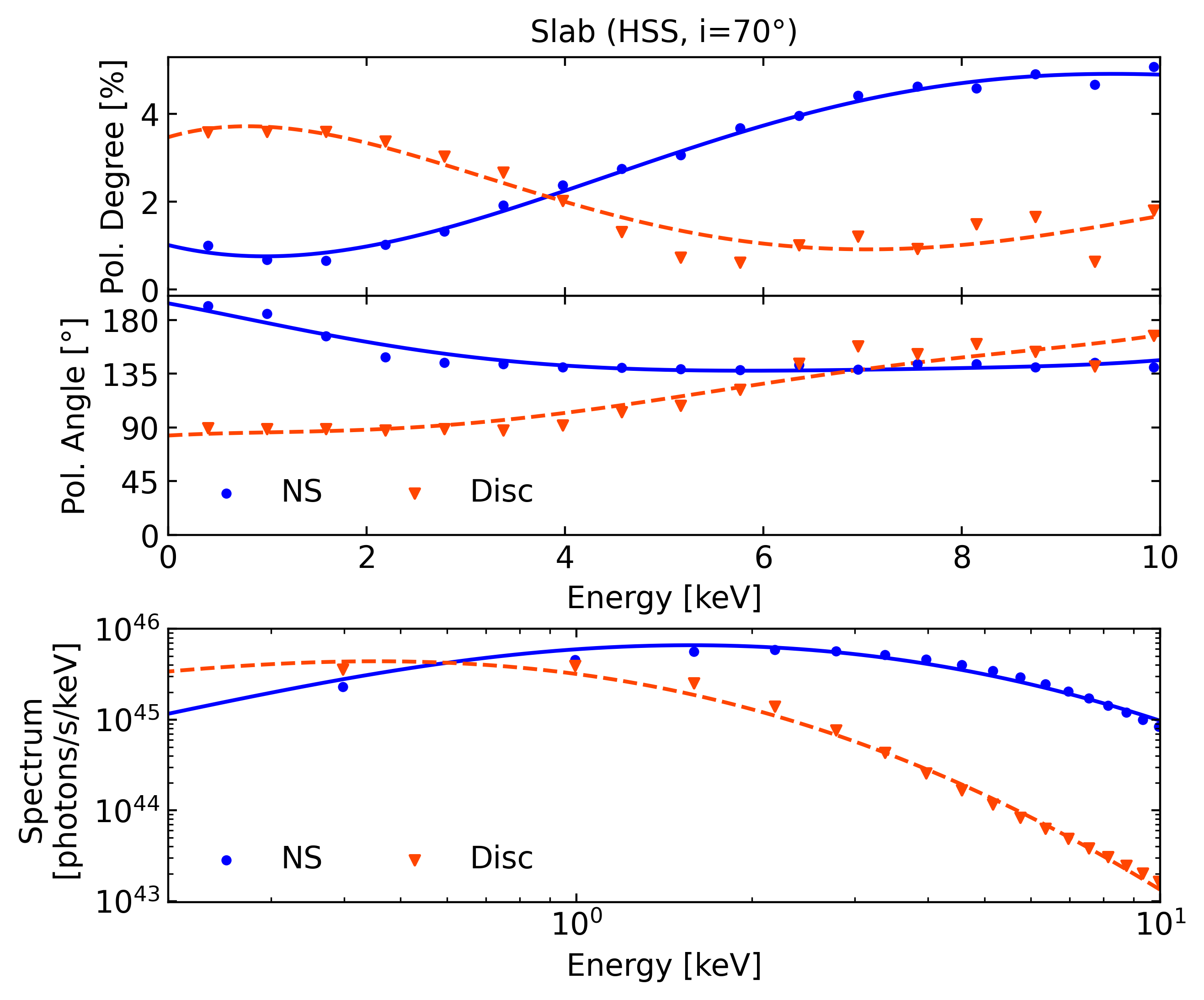}
    \caption{Polarization degree and angle (top panel) and spectrum (lower panel) of NS and disc seed photons (respectively blue dots/solid line and orange triangles/dashed line), including direct radiation coming from the disc, as function of energy between 0.1-10 keV for a LMXB in HSS with a slab corona (see Table \ref{tab:Sim} for all geometrical and physical parameters) observed at $i=70\degree$.}
    \label{fig:Slab.Data}
\end{figure}

The results for the polarization degree and the polarization angle of the two different components, i.e. the NS seed photons scattered by the corona (blue dots and solid line) and those coming from the disc, including both scattered and directly observed photons (orange triangles and dashed line), are shown in Figure \ref{fig:Slab.Data} for an observer at $i=70\degree$. 
Since the thickness of the slab is comparable to the NS radius, most of NS seed photons will hit the corona and eventually get scattered towards the observer, with a significant fraction hitting instead the underlying disc. 
Scattered hot seed photons are expected to be the dominant contribution for energy $\gtrsim 5$ keV with a polarization degree reaching values up to $5\%$.
The polarization angle decreases with energy for the NS component and tends to stabilize around $\approx 150\degree$, while for the disc the polarization angle increases from 90$\degree$ up to $\approx 160\degree$.

In the plot, both the direct Monte Carlo results for the polarization degree and angle, together with the smoothing fit obtained applying a Savitzky-Golay filter \citep{Savitzky.Golay} to the data are shown.
The same will be done in the next figure (relative to the shell geometry), to permit the reader to appreciate the statistical error of the simulations. In all following figures, and for the sake of clarity and illustration, only the smoothing fits will be shown. 
The data noise with respect to the smoothing fits will be very similar to that Figures \ref{fig:Slab.Data} and \ref{fig:Shell.Data}.

Up to now, the disc seed photons have always been assumed to be polarized, as for \cite{Chandrasekhar.1960}. When unpolarized seed photons are considered, the disc radiation directly observed at infinity will not contribute to the total polarization signal. Moreover, the expected polarization degree for the disc contribution will be lower without intrinsic polarization. 

\begin{figure}
    \centering
    \includegraphics[width=0.45\textwidth]{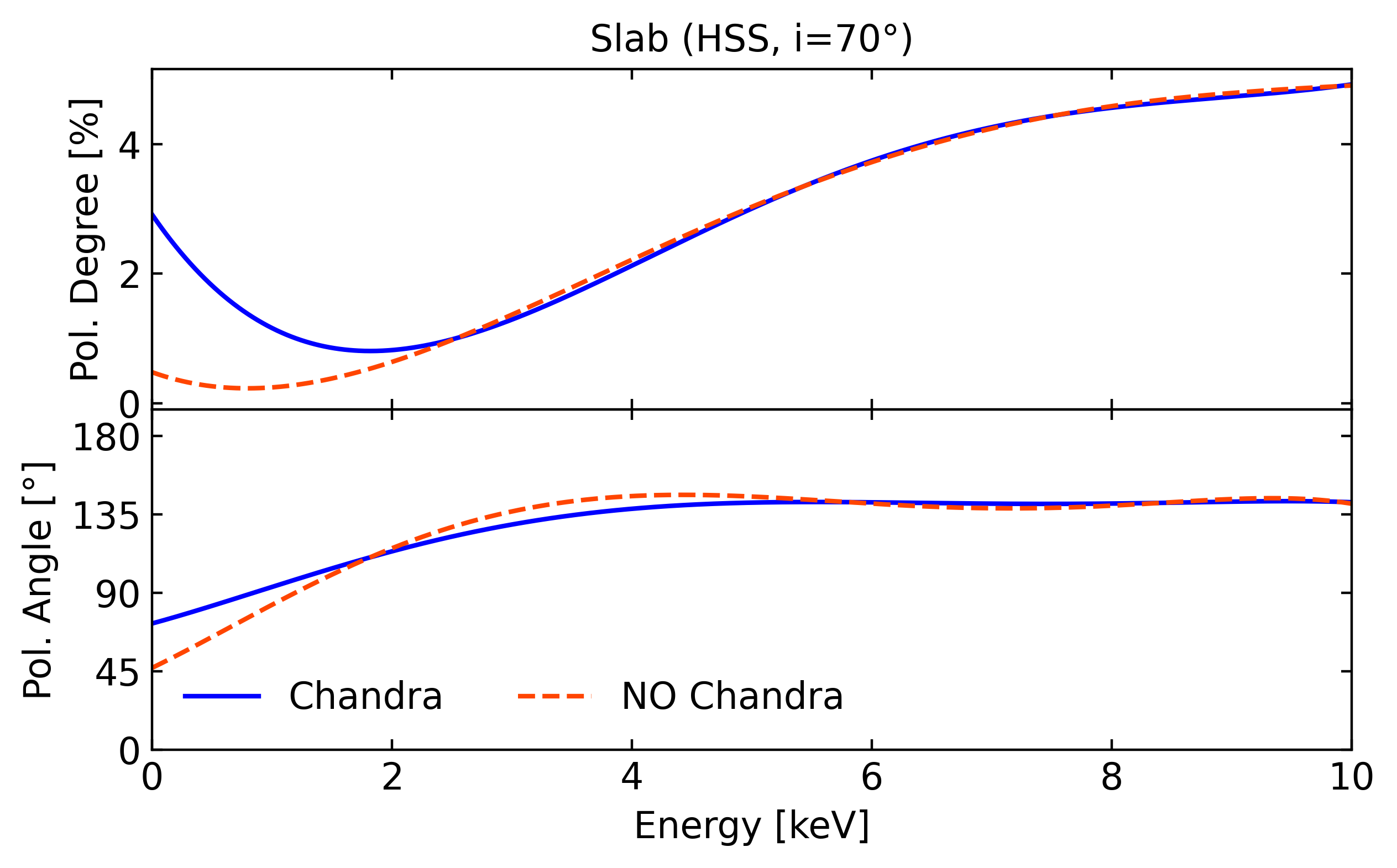}
    \caption{Polarization degree and angle as function of energy between 0.1-10 keV for a LMXB in HSS with a slab corona summing together the NS and disc contributions and
    assuming intrinsic polarization of disc photons (solid blue line) or without (dashed orange line). 
    }
    \label{fig:Slab.Chandra}
\end{figure}

The comparison of the total polarization degree and angle obtained by assuming intrinsic polarized disc photons (blue solid line) or without it (dashed orange line) and summing together the NS and disc components, is illustrated in Figure \ref{fig:Slab.Chandra} for a LMXB in HSS with slab corona observed at $70\degree$. The main differences occur at low energies, where the dominant contribution to the polarization degree comes from the disc photons. At higher ones, instead, the polarization degree behaviour follows the contribution related to scattered NS photons (see also NS component in Figure \ref{fig:Slab.Data}). 

\subsection{Spherical Shell}\label{sec:Shell}

\begin{figure}
    \centering
    \includegraphics[width=0.45\textwidth]{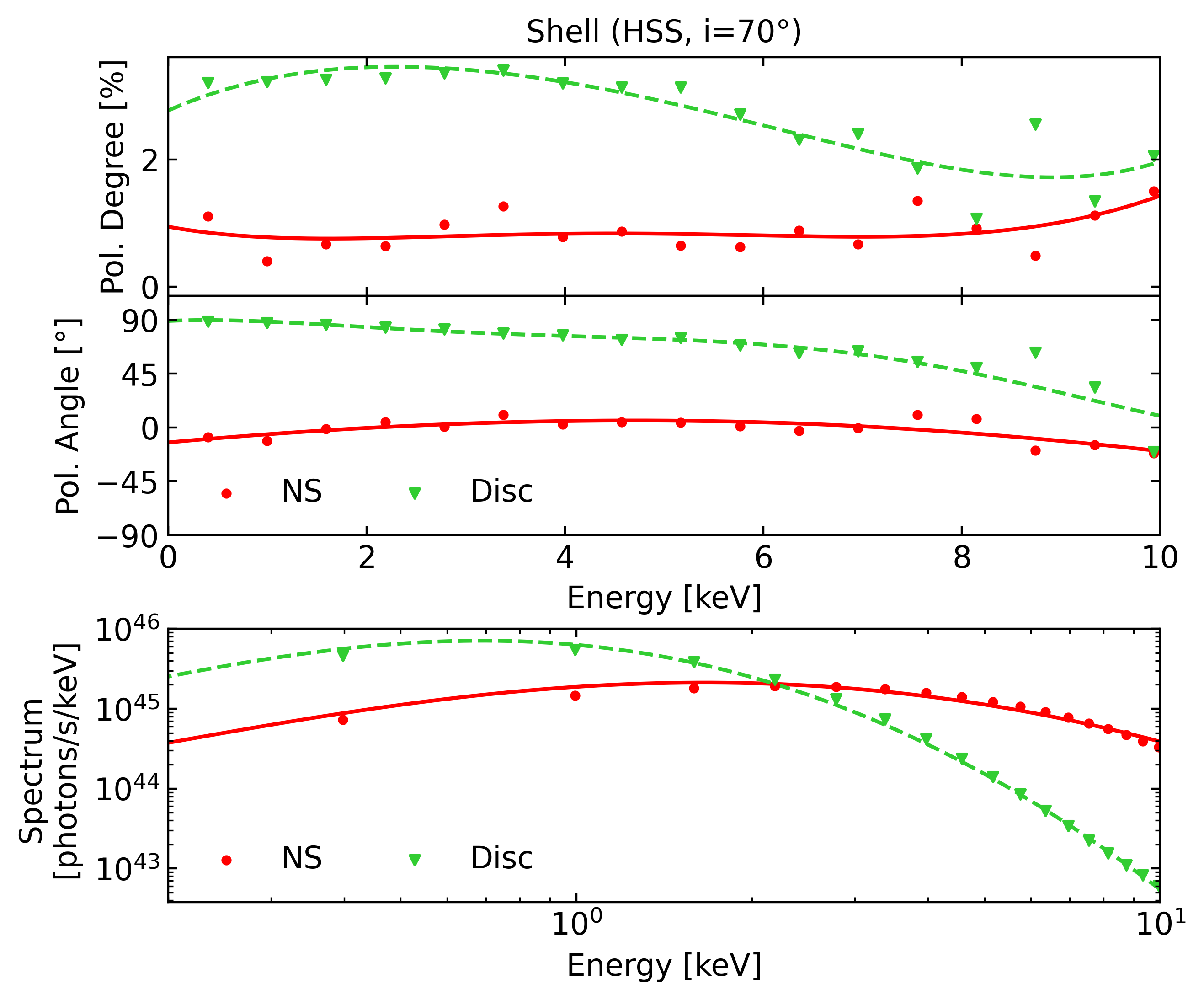}
    \caption{Polarization degree and angle (top panel) and spectrum (lower panel) of NS and disc seed photons (respectively red dots/solid line and green triangles/dashed line), including direct radiation coming from the disc, as function of energy between 0.1-10 keV for a LMXB in HSS with a spherical shell corona (see Table \ref{tab:Sim} for all geometrical and physical parameters) observed at $i=70\degree$.}
    \label{fig:Shell.Data}
\end{figure}

\begin{figure}
    \centering
    \includegraphics[width=0.45\textwidth]{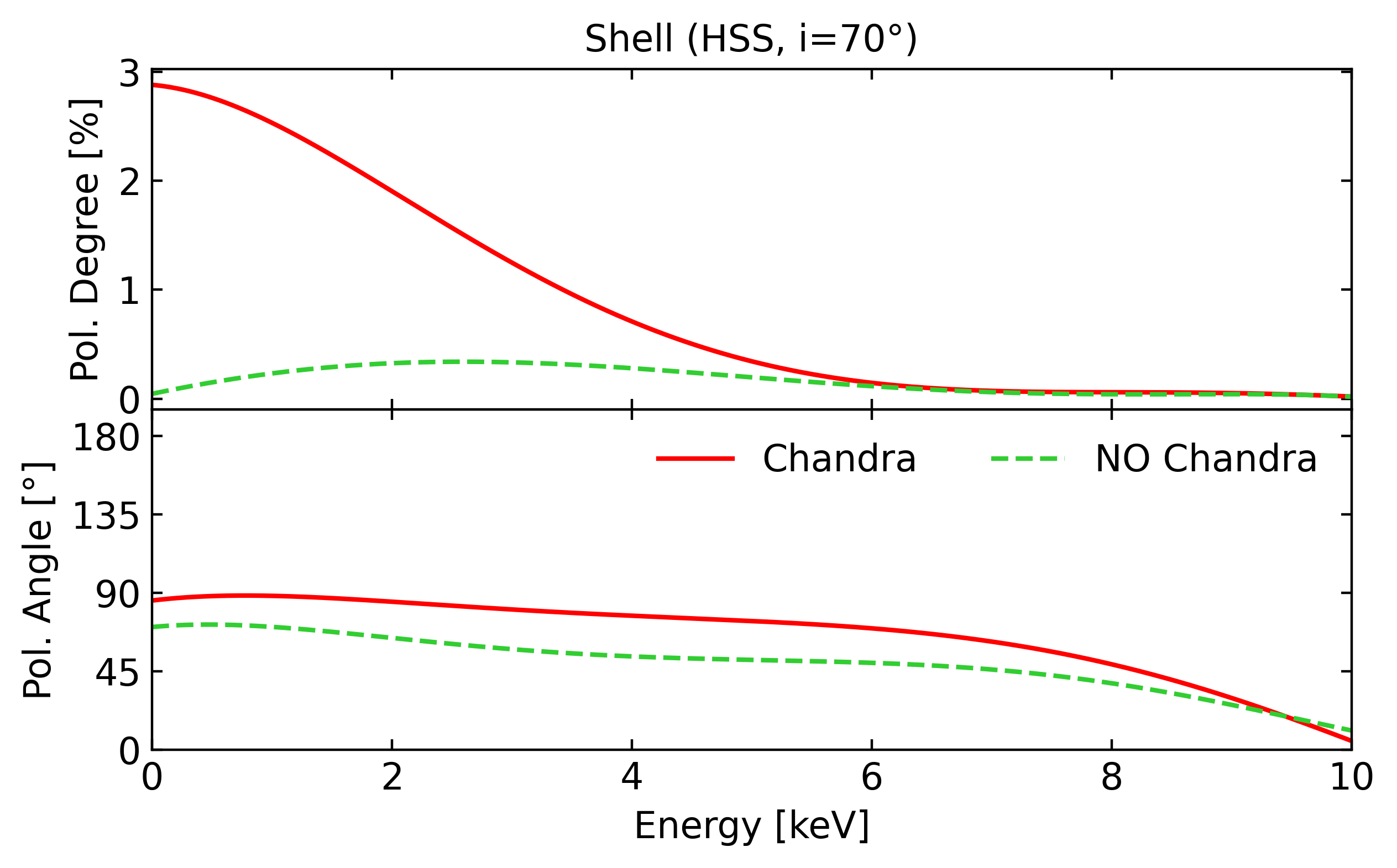}
    \caption{Polarization degree and angle as function of energy between 0.1-10 keV for a LMXB in HSS with a spherical shell corona summing together the NS and disc contributions and
    assuming intrinsic polarization of disc photons (solid red line) or without (dashed green line). 
    }
    \label{fig:Shell.Chandra}
\end{figure}

The spherical shell corona configuration is characterized simply by the inner radius, which can be assumed to be very close to the star, and the outer one. 
Depending on the dimension of the shell, the disc could be partially covered by the corona itself which will therefore intercept mostly the higher energy photons emitted by the inner region of the disc. These photons will be Compton scattered by the corona towards the observer or towards the neutron star or the disc.
For the spherical shell configuration we assume a stationary non-rotating corona \citep{Zhang.etAl.2019}. Since the spin value is not so high, the spacetime metric will be close to the Schwarzschild one and a stationary corona is a quite reasonable assumption.

The temperature of the corona $kT_e$ is equivalent to the one for the slab geometry in order to directly compare the results of the two configurations for the same physical parameters.
The optical depth $\tau$ in spherical geometry is the radial one, with typical value of $\tau_\text{HSS} = 8$ for HSS-LMXBs and $\tau_\text{LHS} = 2$ for LHS-LMXBs.
The physical and geometrical parameters of the corona and the disc are outlined in Table \ref{tab:Sim}.

\begin{figure}
    \centering
    \includegraphics[width=0.45\textwidth]{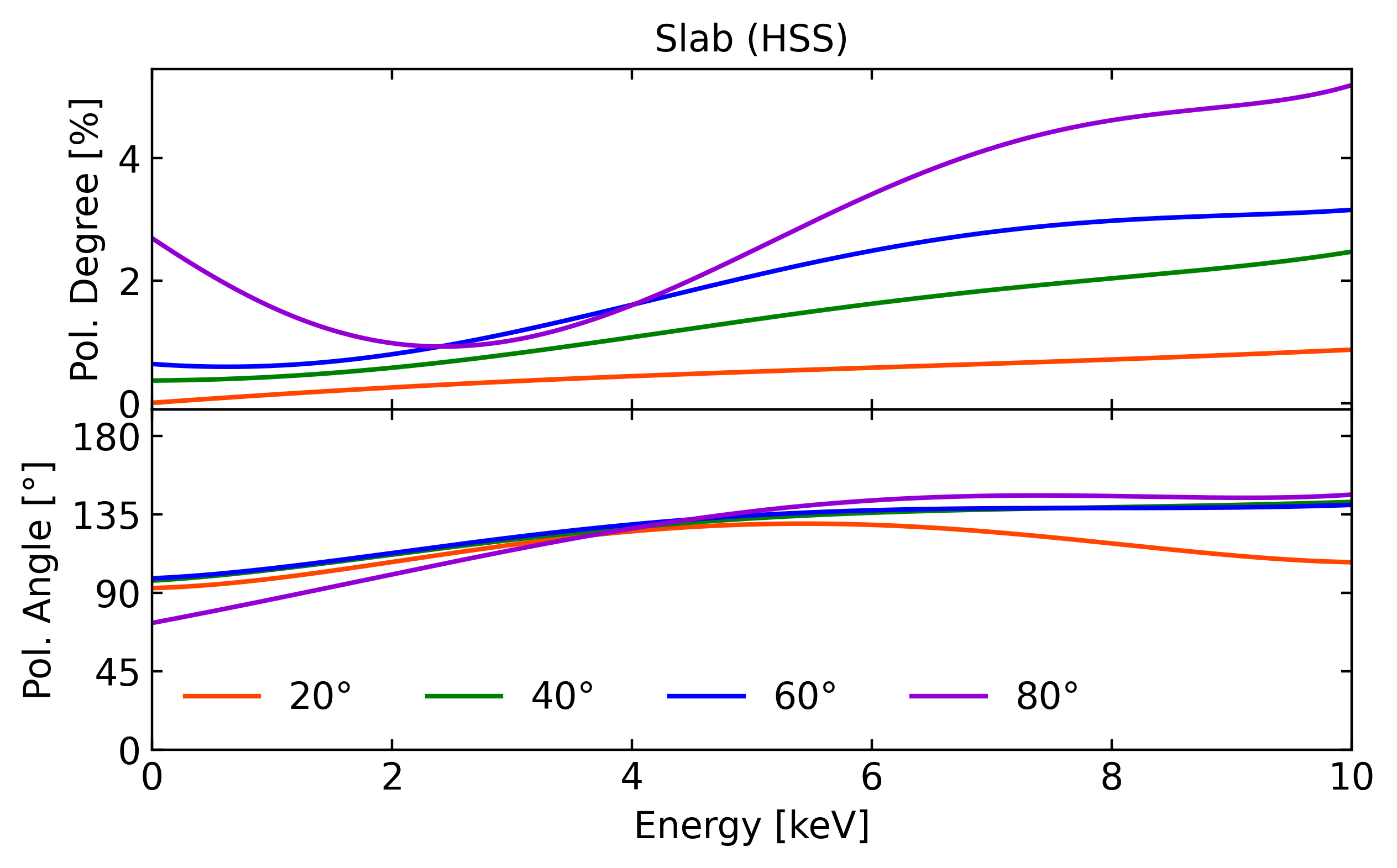}
    \caption{Polarization degree and angle as function of energy between 0.1-10 keV for a LMXB in HSS with a slab corona 
    observed at different inclinations. 
    }
    \label{fig:Slab.Incl}
\end{figure}

\begin{figure}
    \centering
    \includegraphics[width=0.45\textwidth]{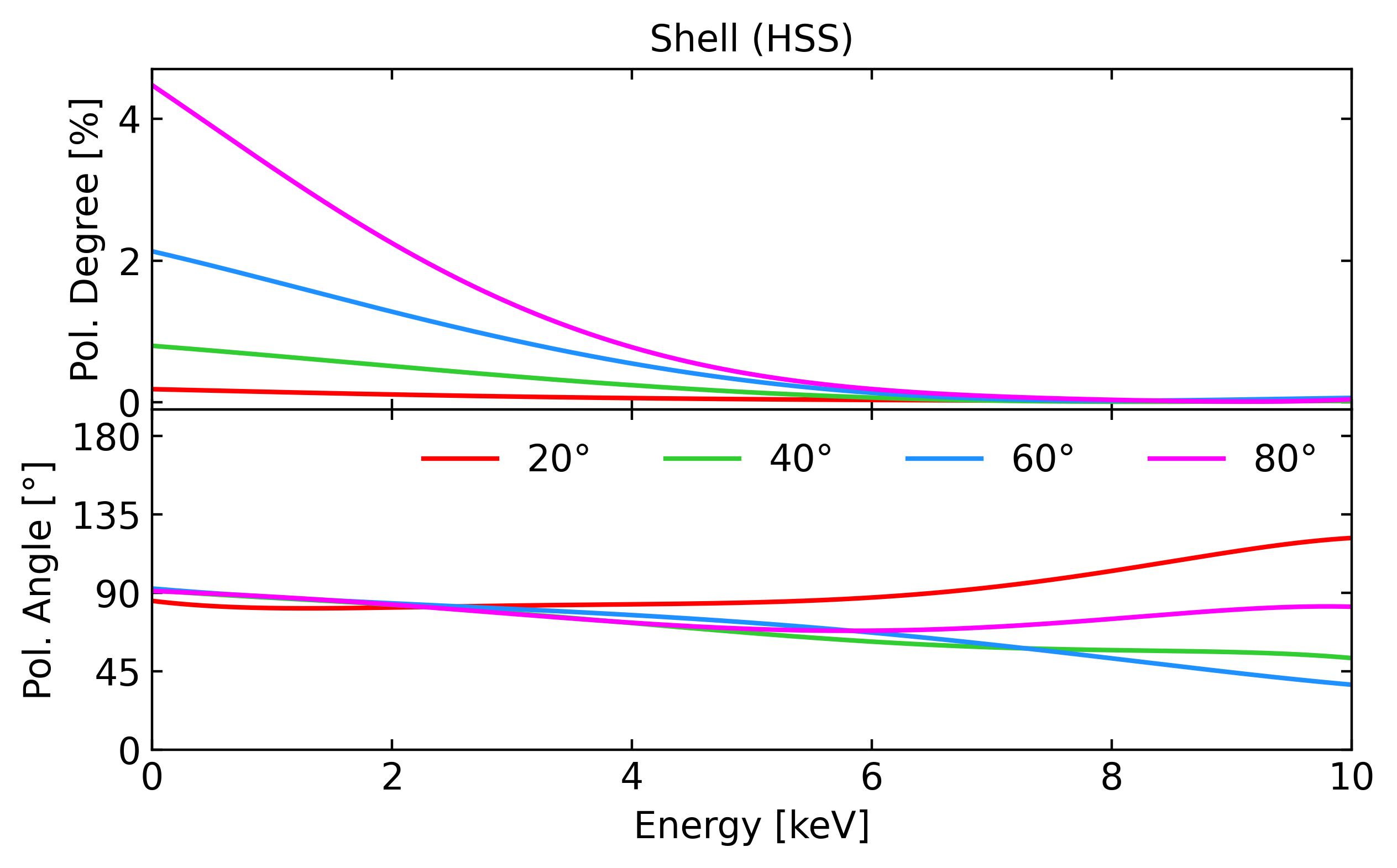}
    \caption{Polarization degree and angle as function of energy between 0.1-10 keV for a LMXB in HSS with a spherical shell corona 
    observed at different inclinations. 
    }
    \label{fig:Shell.Incl}
\end{figure}

The results for the polarization degree and the polarization angle of the two different components, i.e. the NS seed photons scattered by the corona (red dots and solid line) and those coming from th-e disc, including both the scattered and the directly observed photons (green triangles and dashed line), are shown in Figure \ref{fig:Shell.Data} for an observer at $i=70\degree$. 
Since both the NS and the corona are spherically symmetric, the polarization degree related to the NS contribution should be zero. 
The non-zero results are simply artifact of the finite statistic (due to the fact that the polarization degree can assume only positive values), and give an idea of the uncertainty in the simulation (as a sanity check we verified that the Stokes parameters $Q$ and $U$ are oscillating around zero). 
The only contribution to the polarization is therefore due to disc photons directly observed at infinity, since only a small part of the disc is covered by the shell.

Figure \ref{fig:Shell.Chandra} shows the comparison between results assuming intrinsic polarized disc photons (red solid line) and unpolarized ones (dashed green line) for a LMXB in HSS with spherical shell corona observed at $70\degree$, summing together the NS and disc contributions. Also for this corona configuration, the polarization degree at lower energies is about $2\%$ higher considering intrinsic polarized disc photons. 
Moreover, unlike the case with the slab corona, for a shell also the polarization angle significantly changes according to the intrinsic polarization. The polarized disc seed photons (which are the main contribution as direct radiation) have an intrinsic polarization angle of $90\degree$.
Actually, the behaviour of the polarization angle with energy is very close to $90\degree$ for lower energies, where direct and scattered disc photons dominate, and starts decreasing towards low angles (scattered NS photons for a shell configuration exhibit very low polarization degree with a polarization angle very close to $0\degree$, see Figure \ref{fig:Shell.Data}).

\begin{figure}
    \centering
    \includegraphics[width=0.45\textwidth]{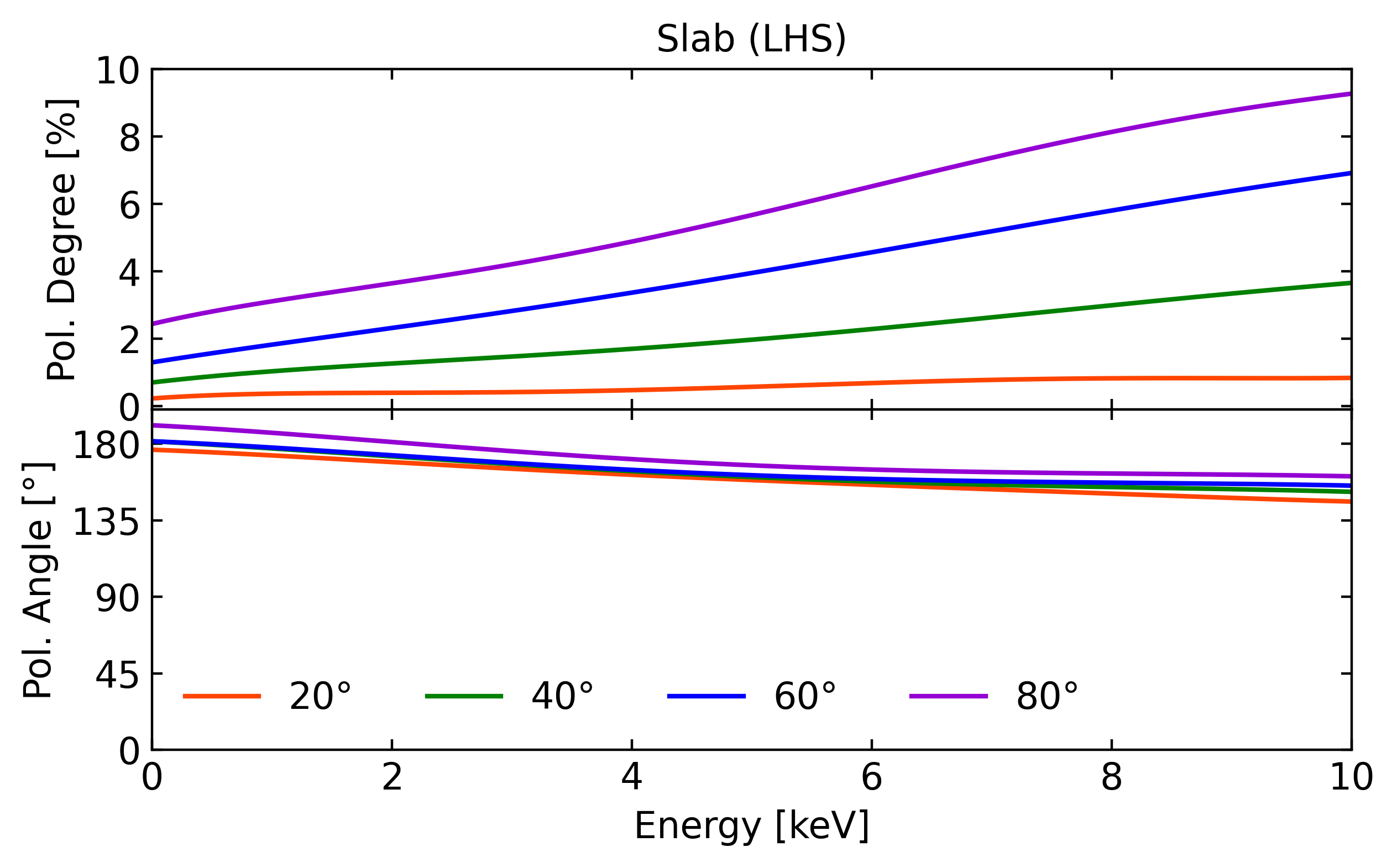}
    \caption{Polarization degree and angle as function of energy between 0.1-10 keV for a LMXB in LHS with a slab corona 
    observed at different inclinations. 
    }
    \label{fig:Slab.LHS.Incl}
\end{figure}

\begin{figure}
    \centering
    \includegraphics[width=0.45\textwidth]{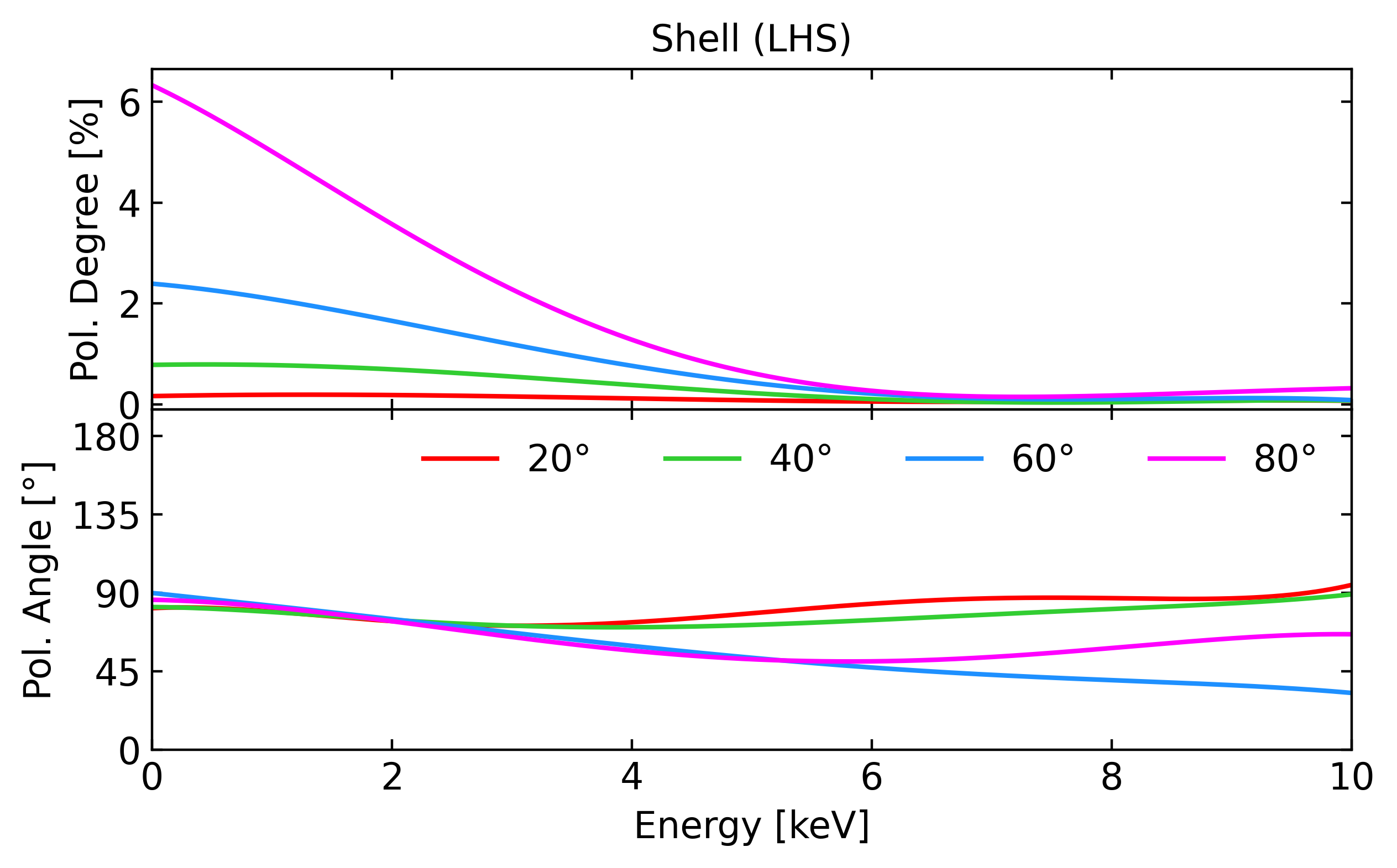}
    \caption{Polarization degree and angle as function of energy between 0.1-10 keV for a LMXB in LHS with a spherical shell corona 
    observed at different inclinations. 
    }
    \label{fig:Shell.LHS.Incl}
\end{figure}


\section{Results}\label{sec:Results}



Figures \ref{fig:Slab.Data} and \ref{fig:Shell.Data} in the previous section showed the results for the NS and the disc contribution separately, for a fixed viewing angle, and only for the HSS. We now discuss how different inclinations affect the polarization properties for both HS and LH states. In the following, results will be presented summing together all the components. 

The polarization degree and angle for different  inclinations ($i=20 \degree$, $40 \degree$, $60 \degree$ and $80 \degree$) for a LMXB in HSS with the slab corona are reported in Figure \ref{fig:Slab.Incl} as a function of energy.
Since the slab thickness is comparable to the NS radius, by looking at the LMXB at a high inclination, the NS surface will be more and more covered by the corona. Photons scattered by the hot electron plasma will produce increasing polarization degree up to $4-5 \%$, with the NS seed components dominating at higher energies and that of the disc at lower ones. 
Decreasing the inclination, a larger region of the disc can be seen by an observer at infinity, together with the direct radiation coming from the NS surface (or boundary layer). The lower the inclination, the lower the polarization degree resulting from the NS and disc contributions ($\lesssim 1 \%$ for $i=20\degree$). 
The polarization angle is almost the same for all the inclinations.

\begin{figure}
    \centering
    \includegraphics[width=0.45\textwidth]{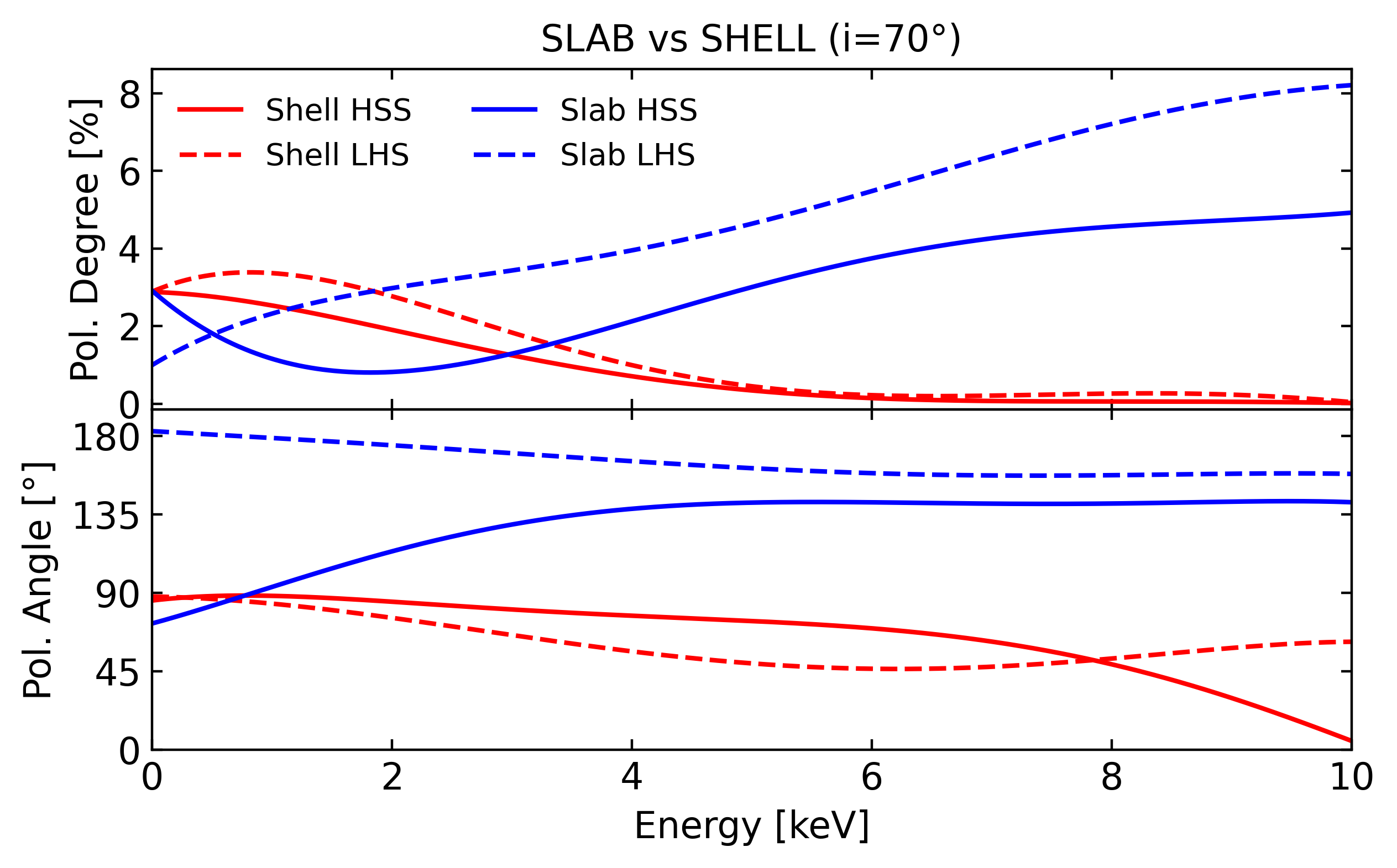}
    \caption{Polarization degree and angle as function of energy between 0.1-10 keV respectively for a LMXB in HSS with a slab (blue solid line) or a spherical shell corona (red solid line) and for a LHS-LMXB with the same coronal geometries, i.e. the slab (blue dashed line) and the spherical shell (red dashed line) observed at $70\degree$ inclination.}
    \label{fig:Diff.Geom}
\end{figure}

Figure \ref{fig:Shell.Incl} shows again the polarization degree and angle for the same inclinations but for a LMXB in HSS with the spherical shell corona.
Also for this configuration, the polarization degree will be significantly less at low inclinations, especially for energies $\lesssim 4-5$ keV.
By looking indeed at the system at low viewing angles, the intrinsic polarization of disc seed photons, which are the dominant contribution to the polarization for the spherical shell configuration, will be significantly lower \citep[see][]{Chandrasekhar.1960}. 
At higher energies, the polarization degree tends to zero for all viewing angles, as expected since scattered NS seed photons become the main contribution and for symmetry reason their total polarization is very close to be null, as appropriate for hemispherical or spherical configurations (see Section \ref{sec:Shell}).

We report also the polarization degree and angle as function of energy for different inclinations for a LMXB in LHS with both slab and shell corona (Figures \ref{fig:Slab.LHS.Incl} and \ref{fig:Shell.LHS.Incl}). 
The overall behaviour for both geometries is very similar to that of HSS case, with higher polarization degree values as the inclination increases. The polarization angle, on the other hand, does not show substantial differences with the inclination.

However, depending on the value of the optical depth of the corona, the full angular dependence of the integrated polarization degree exhibits a drop for both geometries for very high inclinations ($\approx 80\degree - 90\degree$) consistent to the the one found by \cite{Zhang.etAl.2019} and \cite{Ursini.etAl.2022}. 

Since variations in inclinations lead to significant differences in the values of the polarization degree, in principle different geometries (even with the same physical parameters) can present the same behaviour of the polarization degree. Therefore, knowing the inclination (or an upper/lower limit) of a LMXB from the observations, it can be possible to provide a strong constraint on the shape of the corona, with the same fixed physical parameters, by looking at the polarimetric data.

To summarize the results and highlight the difference in the polarization properties between different geometrical models, the polarization for the two different shapes of the corona discussed in this paper are shown in
Figure \ref{fig:Diff.Geom} for a LMXB in HSS (solid lines) and in LHS (dashed lines) for a particular inclination angle ($i=70\degree$). 
The dependence of the polarization properties on the geometry of the corona is clearly evident from the plot. The slab is characterized by an higher polarization degree with respect to the spherical shell geometry, except at low energies where the dominant contribution comes from the disc photons.
The polarization angle is also very different between the two geometries: for the slab the polarization angle increases with the energy, as opposed to what happens for the shell.

Also for a LMXB in LHS, the slab configuration leads to higher polarization degree (up to $8\%$ at higher energies) compared to that obtained for the spherical shell corona (always $\lesssim 4\%$ in the 0.1-10 keV range). In LHS, both coronal geometries are characterized to higher values of the polarization degree with respect to HSS cases. The polarization angle behaviour, moreover, is quite different both for the different geometries and for the two spectral state. 

\section{Conclusions}\label{sec:Conclusions}


In this paper, we presented the first numerical simulations of the X-ray polarization signal expected for weakly magnetized NS-LMXBs, using the fully relativistic Monte Carlo Comptonization code \textsc{monk}. 
We computed the total contribution to the polarization signal considering both scattered NS and disc photons, together with the direct radiation coming from the disc and the star, has been shown for the slab and spherical shell geometries assuming either polarized or unpolarized disc radiation (Figures \ref{fig:Slab.Chandra} and \ref{fig:Shell.Chandra}), and for different viewing inclinations (Figures \ref{fig:Slab.Incl} and \ref{fig:Shell.Incl}).
Higher observation angles and polarized disc photons leads to higher polarization degree for both geometries. 

Both the polarization degree and angle are very distinctive between the two geometries: the slab configuration is characterized by an increasing polarization degree with energy (up to 4-5$\%$ for HSS or 8$\%$ for LHS at higher energies), as opposed to what happens for the spherical shell (Figure \ref{fig:Diff.Geom}).
The polarization angle, moreover, decreases with the energy for the shell configuration (always $\lesssim 90\degree$), while the slab shows higher values up to $\approx 150\degree$ with an increasing/decreasing trends depending on the spectral states.

In summary, the polarization signal in the 2-10 keV band can potentially provide tight constraints on the coronal geometry. For example, observing a low polarization (below 2$\%$) in a low inclination system (below 60$\degree$) would rule out a slab corona, and rather favour a shell geometry. Conversely, a high polarization would be consistent with a slab seen at high inclination. 
Therefore, spectral and polarimetric observations should provide constraints to the geometry of the corona and its physical origin, which is still an open problem.
At the moment, two of the NS-LMXBs in the \ixpe \ first year plan (i.e. GS 1826-238 and Cygnus X-2) have already been observed and a detailed analysis is in progress. 
Novel information on the physics and geometry of these system are being provided by the comparison the data with the calculations presented in this paper.


\section*{Acknowledgements}

SB, AG, GM and FU acknowledge financial support from ASI (grant 2017-12-H.0).
WZ acknowledges the support by the Strategic Pioneer Program on Space Science, Chinese Academy of Sciences through grant XDA15052100.
We thank the Referee for the helpful comments that have helped to improve the quality of the manuscript.


\section*{Data Availability}

The code underlying this article, \textsc{monk}, is proprietary. Simulation data supporting the findings of the article will be shared on reasonable request.



\bibliographystyle{mnras}
\bibliography{references.bib}







\bsp	
\label{lastpage}
\end{document}